\documentclass{article}
\usepackage{PRIMEarxiv}

\usepackage[inline]{enumitem}
\usepackage{pbox}
\usepackage{balance}
\newlist{inlinelist}{enumerate*}{1}
\setlist*[inlinelist,1]{label=\roman*),itemjoin={{, }},itemjoin*={{, and }}}
\usepackage{url}
\usepackage[normalem]{ulem}
\usepackage{multirow}
\usepackage{multicol}
\usepackage{booktabs}
\usepackage{amsfonts}
\usepackage{amsmath}
\usepackage{amssymb}
\usepackage{amsthm}
\usepackage{adjustbox}
 \usepackage[normalem]{ulem}

\title{Extending English IR methods to multi-lingual IR}

\author{
     Carlos Lassance \\
  Naver Labs Europe \\
  France\\
  \texttt{carlos.lassance@naverlabs.com}
}

\begin{document}

\maketitle


\begin{abstract}
This paper describes our participation in the 2023 WSDM CUP - MIRACL challenge. Via a combination of i) document translation; ii) multilingual SPLADE and Contriever; and iii) multilingual RankT5 and many other models, we were able to get first place in both the known and surprise languages tracks. Our strategy mostly revolved around getting the most diverse runs for the first stage and then throwing all possible reranking techniques. While this was not a first for many techniques, we had some things that we believe were never tried before, for example, we train the first SPLADE model that is effectively capable of working in more than 10 languages. However, a more careful study of the results is needed in order to verify if we were able to get first place just due to brute force or if the hybrids we developed really brought improvements over the other team's solutions.
\end{abstract}

%
\maketitle

\section{Introduction}
In this paper, we detail our WSDM CUP 2023 - MIRACL submissions. Due to the time crunch we were during the competition, this will mostly read as an incremental improvement report, detailing which steps we took and how it evolved. This will differ vastly from a traditional research paper where models are trained and evaluated to be comparable, here they will mostly build on top of one another as we were developing for the competition. Compared to other strategies, what we believe were the three main pillars of ours and that made a difference compared to other teams were: \begin{inlinelist}
    \item the use of NLLB-200 for the translation of the corpuses back to English, allowing us to use the models we already had in English; 
    \item the training of a novel SPLADE model capable of working in the original 16 languages of the MIRACL dataset; 
    \item Training the first multilingual RankT5~\cite{rankt5}.
\end{inlinelist}. We aim to make all these models available and as easy to reproduce as we can for the community, but this will unfortunately still take quite some time to be possible. 

We follow the strategy we used on our latest TREC notebooks, in that we strive for making this more streamlined than a normal research paper would be.  We will now present a list of the papers that better introduce and detail the models we used here and refer the reader to check them for a better explanation than those we have here, that are mainly dedicated to how to apply it to MIRACL and not to the methods themselves: \begin{inlinelist}
    \item Training non English SPLADE models~\cite{scratch} 
    \item The SPLADE model~\cite{pp,efficiency}
    \item The Contriever model and its pretraining~\cite{contriever}
    \item The RankT5 reranker~\cite{rankt5}
    \item MonoT5~\cite{monoT5}
    \item The LCE loss~\cite{rerankers}
    \item ColBERT~\cite{colbert}
    \item For our ensembling we use Ranx~\cite{ranx} and their min-max normalized sum ensembling
\end{inlinelist}.

\subsection{The MIRACL dataset}

The MIRACL dataset~\cite{miracl} is composed of 16 languages and relies on monolingual retrieval. Those languages are were we spent most of our time and will detail our approaches in Sections 2 and 3. Two languages are added at the end as the ``surprise'' languages, without training data, to which we use Section 4 to detail how we generated our runs. Finally, we do a quick analysis on queries and labelling in Section 5. 

\section{Training the first stage}
\label{sec:first_stage}

For the first stage of retrieval, we report step by step how our models evolved during the competition. We note that some editing of the history was done in order to make it more concise, but this is pretty much the order in which things evolved during the two months of MIRACL. We detail our results in the dev set, with incrementing tables of results, leading to the final hybrid that will be used for reranking.

\subsection{The initial baseline}

We start with the baseline provided by the organizers which include an mDPR model trained on mMARCO and BM25. We ensemble these models and generate \emph{HYBRID 0}, which we present in Table~\ref{tab:initial_baseline}. We already start noticing that ensembling is strong in this dataset, as the ensemble has almost $1.4\times$ the effectiveness (nDCG@10) of mDPR which was the better model in the ensemble and the recall@100 is already close to 90\%.

\begin{table}[ht]
\centering
\caption{Baselines provided by the organizers. Just a simple hybrid is a good baseline.}
\label{tab:initial_baseline}
\adjustbox{max width=\columnwidth}{%
\begin{tabular}{cc|ccc}
\toprule
\# & model                    & nDCG@10 & Recall@20 & Recall@100 \\
\midrule
\multicolumn{5}{c}{Models} \\
\midrule
a & BM25                & 39.3    & 60.9\% & 78.7\%      \\
b & mDPR                & \textbf{41.5}    & \textbf{62.8\%} & \textbf{78.8\%}      \\
\midrule
\multicolumn{5}{c}{Hybrids} \\
\midrule
0 & a+b                 & 57.8    & 83.0\% & 93.7\%     \\
\bottomrule
\end{tabular}
}
\end{table}

\subsection{Going back to English leads to improvement}

The next step was to translate the corpora into English. This is the simplest way of working in any language, as English translators are pretty strong and there is a multitude of retrievers and rerankers available off-the-shelf for English data. It also adds the knowledge of the translator into the mix. Ideally, we would have used something like google translate or deepL that are already effective and resilient to problems, but due to a lack of a free-to-use API\footnote{Some solutions exist, but we were not sure if we could actually use those as they mostly break the terms of the APIs.}. Thus I've used the NLLB200 model~\cite{nllb}, which \begin{inlinelist} \item returned a lot of weird stuff that we filtered manually (things like "I don't know it" or "I won't translate this" came out). \item was too expensive to run with the largest version, thus for the small corpora we used the large NLLB (3B params) and for the corpora larger than 4M documents we used the distilled version \item Due to the cost of the models we only translated the first 128 tokens. \end{inlinelist} From initial experiments we have noted a very sharp drop in performance using the distilled version, but it would be too costly to use the other for the largest datasets. 

Now that every corpus is translated to English, we took one of the SPLADE++~\cite{pp} models and fine-tuned 16 different versions one on each translation (in English we just use the MIRACL corpus). This led to what we call T-SPLADE, which added to the BM25 and mDPR leads to ``HYBRID 1''. We noticed a large gain from adding the T-SPLADE, even individually it is not stronger ``than HYBRID 0''. One thing of note is that just adding these models makes the Recall@20 achieve 90\% and Recall@100 already surpass 97\%, meaning that improving over this will be hard.

\begin{table}[ht]
\centering
\caption{Result comparison adding translated-SPLADE.}
\label{tab:add_translate}
\adjustbox{max width=\columnwidth}{%
\begin{tabular}{cc|ccc}
\toprule
\# & model                    & nDCG@10 & Recall@20 & Recall@100 \\
\midrule
\multicolumn{5}{c}{Models} \\
\midrule
a & BM25                & 39.3    & 60.9\% & 78.7\%      \\
b & mDPR                & 41.5    & 62.8\% & 78.8\%      \\
c & T-SPLADE            & \textbf{54.5}    & \textbf{71.3\%} & \textbf{83.3\%}      \\
\midrule
\multicolumn{5}{c}{Hybrids} \\
\midrule
0 & a+b                 & 57.8    & 83.0\% & 93.7\%     \\
1 & a+b+c               & 70.0    & 90.4\% & 97.2\%     \\
\bottomrule
\end{tabular}
}
\end{table}

\subsection{Adding monolingual SPLADE for non-English languages}

Due to recent submissions~\cite{scratch} and the TREC-NeuCLIR 2022 we had already 5 pretrained SPLADE in Arabic, Chinese, Japanese, Persian, and Russian. Of those we noticed that the Arabic and Chinese did not generalize well to MIRACL, so we just used the other three. Adding these models to the fold increased effectiveness by a small margin, but not enough to be considered a new baseline. Actually, contrary to what we had seen in other datasets, training monolingual SPLADE did not work that well in MIRACL, compared to multilingual, which we confirm in the following paragraphs.

\subsection{The introduction of multi-lingual SPLADE - you have to fail first to make it work}

Now to the only possibly novel part of our submission, multi-lingual SPLADE models (mSPLADE). This has already kinda been tried in SPLADE-X, which was not multi-lingual but cross-lingual, where they had troubles: not only with the tokenizer size (which vastly increases memory use), but also with the pretrained models. For the pre-trained models we had the solution, pre-train from scratch on MIRACL following~\cite{scratch}, however tokenizer size, and the amount of memory we need as it increases we had no solution. Actually, we got lucky that we had access to a few A100's with 80g which are the only ones that could train models that size and we are limited to training with at most 128 tokens and indexing with at most 256.

For the tokenizer, we initially used the tokenizer from XLM-Roberta, which has 240k+ tokens, and then pretrained with MLM+FLOPS on the 15 non-English corpora of MIRACL. We then finetuned first on the languages of mMARCO that are present on the 15 (Arabic, Chinese, French, Hindi, Indonesian, Japanese, Russian, Spanish) and finally finetuned on MIRACL using two different values of lambda (mSPLADE-small and mSPLADE-large for the length of the documents). However, while the effectiveness was better than the translated models we were not satisfied. Note that as these models are not trained for English, we consider that they have the same effectiveness as the best of mDPR and BM25 when averaging.

Thus, we went back and recreated a tokenizer from scratch with just 120k tokens, but based on the statistics of the 15 non-english MIRACL datasets. We then pre-trained with MLM+FLOPS on MIRACL, finetuned on mMARCO and finally finetuned on MIRACL to get a new model that we call mSPLADE-sTok. Note that we still used the same max sequence length of 128 for training, but it was way faster to redo all steps for this new model. With more patience, i.e. increasing pretraining steps and the max sequence lengths these models would probably be better. mSPLADE-sTok is not only faster to run, but it is also much more stable and effective compared to mSPLADE-small and mSPLADE-large. Finally, as this was a competition we added all three models to the ensemble, creating ``HYBRID 2'', which improves upon the previous one, achieving a recall@20 that is almost as good as the recall@100 of our first hybrid. We note that there could probably be even more gains by properly tuning the hyperparameters (sTok has smallish docs and large queries, which is not ideal for SPLADE) and by letting it train for more time, especially the MLM+FLOPS pretraining part, which uses only 5 epochs

\begin{table}[ht]
\centering
\caption{Result comparison by adding mSPLADE models.}
\label{tab:add_msplade}
\adjustbox{max width=\columnwidth}{%
\begin{tabular}{cc|ccc}
\toprule
\# & model                   & nDCG@10 & Recall@20 & Recall@100 \\
\midrule
\multicolumn{5}{c}{Models} \\
\midrule
a & BM25                & 39.3    & 60.9\% & 78.7\%      \\
b & mDPR                & 41.5    & 62.8\% & 78.8\%      \\
c & T-SPLADE            & 54.5    & 71.3\% & 83.3\%      \\
d & SPLADE-mono         & \multicolumn{3}{c}{Only available in 3 languages} \\
e & mSPLADE-small       & 57.9    & 76.2\% & 88.7\%      \\
e & mSPLADE-large       & 60.2    & 78.6\% & 90.6\%      \\
f & mSPLADE-sTok        & \textbf{63.9}    & \textbf{81.6\%} & \textbf{92.4\%}      \\
\midrule
\multicolumn{5}{c}{Hybrids} \\
\midrule
0 & a+b                 & 57.8    & 83.0\% & 93.7\%     \\
1 & a+b+c               & 70.0    & 90.4\% & 97.2\%     \\
2 & a+b+c+d+e+f         & 74.2    & 92.1\% & 98.3\%     \\
\bottomrule
\end{tabular}
}
\end{table}

\subsection{Proper dense multilingual retrieval - mColbert and mContriever enter the fold}

In the previous paragraphs, we did our best to improve sparse multilingual retrieval, but we have no base to compare it to dense models trained on MIRACL. Thus we now add mContriever and mContriever-ColBERT. The idea is simple, we take the mContriever finetuned on mMARCO and further finetune it on MIRACL. To generate mColBERT we take the mContriever we just finetuned and further train it on MIRACL, but this time under the ColBERT-brute force framework. If we disregard English (for which mSPLADE is not trained) mSPLADE-sTok has similar effectiveness to both, and by adding all models together ``HYBRID 3'' we get a very impressive first stage ranker. However, we are starting to get diminishing returns, especially considering that the models we are adding are much better than the initial ones.

\subsection{Learning to ensemble}

Considering the diminishing returns we had with adding the models we consider two more hybrids: \begin{itemize}
    \item ``HYBRID BEST'' is learned by testing all the ensembling possibilities and keeping the one with the best nDCG@10 on the dev set
    \item ``HYBRID SIMPLE'' tries to simplify, keeping only a model per category, thus: \begin{inlinelist}
        \item lexical retrieval: BM25
        \item translated retrieval: T-SPLADE
        \item sparse multilingual retrieval: mSPLADE-sTok
        \item dense multilingual retrieval: mContriever.
    \end{inlinelist}
The goal of this hybrid is to check how much we lose by not going brute-force with all models. \end{itemize}. We display the final results in Table~\ref{tab:all_first}. We note that by learning the best hybrid we can get a little bit more improvement, achieving a better recall@20 than the baseline hybrid achieved with recall@100. The SIMPLE hybrid with just 4 models is capable of almost the same performance as hybrid 2 which uses 7 different methods. Finally, we still have to verify if everything we did in the dev set generalizes to one of the test sets.

\begin{table}[ht]
\centering
\caption{Comparison of all our first stage models on the dev set.}
\label{tab:all_first}
\adjustbox{max width=\columnwidth}{%
\begin{tabular}{cc|ccc}
\toprule
\# & model                    & nDCG@10 & Recall@20 & Recall@100 \\
\midrule
\multicolumn{5}{c}{Models} \\
\midrule
a & BM25                & 39.3    & 60.9\% & 78.7\%      \\
b & mDPR                & 41.5    & 62.8\% & 78.8\%      \\
c & T-SPLADE            & 54.5    & 71.3\% & 83.3\%      \\
d & SPLADE-mono         & \multicolumn{3}{c}{Only available in 3 languages} \\
e & mSPLADE-small       & 57.9    & 76.2\% & 88.7\%      \\
f & mSPLADE-large       & 60.2    & 78.6\% & 90.6\%      \\
g & mSPLADE-sTok        & 63.9    & 81.6\% & 92.4\%      \\
h & mContriever         & 64.6    & 82.7\% & 93.2\%      \\
i & mColBERT            & \textbf{65.5}    & \textbf{83.5\%} & \textbf{93.6\%}      \\
\midrule
\multicolumn{5}{c}{Hybrids} \\
\midrule
0 & a+b                            & 57.8    & 83.0\% & 93.7\%     \\
1 & a+b+c                          & 70.0    & 90.4\% & 97.2\%     \\
2 & a+b+c+d+e+f+g                  & 74.2    & 92.1\% & 98.3\%     \\
3 & a+b+c+d+e+f+g+h+i              & 75.9    & 93.1\% & 98.8\%     \\
Best & $\max$(a+b+c+d+e+f+g+h+i)   & \textbf{77.2}    & \textbf{94.0\%} & \textbf{98.9\%}     \\
Simple & a+c+f+g                   & 74.0    & 91.8\% & 98.1\%     \\
\bottomrule
\end{tabular}
}
\end{table}

\section{Training the rerankers}
\label{sec:rerank}

Now that we finished with the first stage retrievers is time for the last effectiveness increment: reranking. We note that due to the already high effectiveness of the first stage, it is not clear which top$k$ we should rerank, as even if recall@k+i is larger than recall@k, adding more documents always adds noise to reranking, so we always test @\{10,20,100\} and present the best of the three. All of our rerankers are trained using a set of negatives coming from the top1k of our best model at the time (which makes it harder to reproduce). We aim at making these models available as we have done for TREC-DL and TREC-neuCLIR, but it could take a while to actually have them online.

\subsection{monoMT5 and RankT5}

To set up our main rerankers, we start from the mono MT5-13b trained on MMARCO that got impressive results on TREC neuCLIR 22 by the Unicamp's team\footnote{available at: unicamp-dl/mt5-13b-mmarco-100k}. We then first finetune that model following the monoT5 recipe, with only the negatives from the official MIRACL qrels. We then also finetune it as a RankT5-Encoder only~\cite{rankt5}\footnote{We use RankT5-Encoder only as it has very similar results to full RankT5, but it is much less costly as it removes the decoder}, this time using the negatives from the best model we had at the time. 

We present the results of reranking ``Hybrid best'' in Table~\ref{tab:first_rerankers}. We notice that mono MT5 does not improve over the first stage hybrid, kinda showing that we probably could have trained it better, however when we look into RankMT5 it actually improves slightly over the first stage hybrid, and combining the first stage hybrid with the two rerankers (Simple Reranking Hybrid) allows us to reach an nDCG@10 of 80.4 that would be the best in the public dev leaderboard.\footnote{We note that a lot of teams stopped submitting to the public dev, using the leaderboard solely for test-a, thus making it hard to compare on the dev only.}.

\begin{table}[ht]
\centering
\caption{Reranking with mT5-13b.}
\label{tab:first_rerankers}
\adjustbox{max width=\columnwidth}{%
\begin{tabular}{cc|cc}
\toprule
\# & model                    & nDCG@10 & \# in simple reranking hybrid \\
\midrule
\multicolumn{4}{c}{First stage hybrids} \\
\midrule
\multicolumn{2}{c|}{Best} & 77.2    & 16     \\
\midrule
\multicolumn{4}{c}{Rerankers} \\
\midrule
$\alpha$ & mono mT5-13b & 76.9    &  14    \\
$\beta$  & Rank mT5-13b & 78.6    &  15     \\
\midrule
\multicolumn{4}{c}{Reranking hybrids} \\
\midrule
Simple reranker hybrid & max(Best first stage + $\alpha$ + $\beta$) & 80.4 & n/a     \\
\midrule
\bottomrule
\end{tabular}
}
\end{table}

\subsection{Throwing everything but the kitchen sink}

We were surprised by the lack of improvement from the rerankers we used, so we started testing a lot of different rerankers. Finally, we ended up with 6 rerankers, the 2 we presented in the previous subsection and an extra 4 RankT5-Encoder-only models, all trained in a similar fashion, but starting from different pretrained language models, namely: Bloomz-3b~\cite{bloomz}\footnote{\url{bigscience/bloomz-3b}}, mT0(xl\footnote{\url{https://huggingface.co/bigscience/mt0-xl}} and xxl\footnote{\url{https://huggingface.co/bigscience/mt0-xxl}})~\cite{bloomz} and a flan-T5-xl~\cite{flant5}\footnote{\url{https://huggingface.co/google/flan-t5-xl}} trained solely in English that we run over the translated corpora. 

\paragraph{English special case}

Considering that we already had a bunch of rerankers finetuned for English due to our participation in TREC-DL 2022, we tested a few of them and actually noticed a slight improvement (72.9 nDCG@10 improves to 73.5 nDCG@10) when we added the monoT0pp reranker to the hybrid. In other words the actual reranker ensemble considers 6 rerankers for all languages, but on English, we actually consider 7.

\paragraph{Results on evaluation set}

We present the results of all rerankers over ``Hybrid best'' in Table~\ref{tab:final_reranking}. We notice that none of our rerankers outperformed the first stage by more than 1.5 nDCG@10 and that adding these 4 rerankers improves our hybrid from 80.4 to 81.0 which was the best result on the dev set, but it is not that far from what we could get with just our first stage retrievers (77.2). In other words, we probably need to get a better way of training rerankers (same conclusion as our TREC-DL paper). We also note that the first stage was the most important part, appearing in all languages, while BloomZ was actually only useful for Arabic and Bengali.

\begin{table}[ht]
\centering
\caption{Full reranking results over ``Best'' first stage hybrid.}
\label{tab:final_reranking}
\adjustbox{max width=\columnwidth}{%
\begin{tabular}{cc|ccc}
\toprule
\# & model                   & nDCG@10 & \# in simple rr hybrid & \# in best rr hybrid \\
\midrule
\multicolumn{5}{c}{First stage hybrids} \\
\midrule
F & Best & 77.2    & 16 &  16   \\
\midrule
\multicolumn{5}{c}{Rerankers} \\
\midrule
$\alpha$   & mono mT5-13b & 76.9    &  14 & 12   \\
$\beta$    & Rank mT5-13b & 78.6    &  15 & 13   \\
$\gamma$   & Rank mT0-xl & 77.8    &  15 & 10   \\
$\delta$   & Rank mT0-xxl & 78.2    &  15 & 8   \\
$\epsilon$ & Rank Bloomz & 71.2    &  15 & 2   \\
$\zeta$    & Translate + Rank T5 & 70.0    &  15 & 4   \\
\midrule
\multicolumn{5}{c}{Reranking hybrids} \\
\midrule
Simple & max(F + $\alpha$ + $\beta$) & 80.4 & -     \\
Best & max(F + $\alpha$ + $\beta$ + $\gamma$ + $\delta$ + $\epsilon$ + $\zeta$) & 81.0 & -     \\
\midrule
\bottomrule
\end{tabular}
}
\end{table}

\paragraph{Results on the test sets}

Now that we have described our approach and how it evolved over the development set, we can finally talk about the leaderboard and the competition itself, including two test sets, A and B. Test set A was released first, with a 12-language subset (removing 4 languages, Spanish, Farsi, French, and Chinese). Test set B was the final evaluation set, and was released in the final days of the challenge. Unfortunately, we did not keep track of our different hybrids and rerankers during the challenge period, thus we can only evaluate the final submission.

We present the results for the top 3 teams on test-A in Table~\ref{tab:testa-leaderboard} and for test-B in Table~\ref{tab:testb-leaderboard}.

\begin{table}[ht]
\centering
\caption{Test-a leaderboard results.}
\label{tab:testa-leaderboard}
\adjustbox{max width=\columnwidth}{%
\begin{tabular}{cc|c|cccccccccccccccc}
\toprule
\# & team       & Average       & ar            & bn            & en            & fi            & id            & ja            & ko            & ru          & sw            & te            & th            \\
\midrule
1  & nle \textbf{(ours)} & \textbf{81.2} & \textbf{84.2} & 91            & \textbf{64.5} & \textbf{78.6} & \textbf{62.9} & \textbf{83.4} & 77.7          & \textbf{83} & \textbf{86.6} & \textbf{93.8} & 87.6          \\
2  & ew         & 81            & 83.1          & \textbf{92.2} & 66.3          & 76.9          & 61.2          & 83            & \textbf{80.1} & 81.7        & 85.3          & 93.1          & \textbf{87.9} \\
3  & vector     & 80.7          & 83.1          & 91.6          & 62.4          & 77.3          & \textbf{62.9} & 82.2          & 78.1          & 82.4        & 86.8          & 93.6          & 87.5    \\    
\bottomrule
\end{tabular}
}
\end{table}

\begin{table}[ht]
\centering
\caption{Test-b leaderboard results.}
\label{tab:testb-leaderboard}
\adjustbox{max width=\columnwidth}{%
\begin{tabular}{cc|c|cccccccccccccccc}
\toprule
  &          & Average       & ar            & bn            & en            & es            & fa            & fi            & fr            & hi            & id            & ja            & ko            & ru            & sw            & te            & th            & zh            \\
\midrule
1 & nle \textbf{(ours)}      & \textbf{73.3} & \textbf{72.8} & 81.6          & \textbf{83.1} & 74.6          & \textbf{72.1} & \textbf{69.2} & \textbf{71.8} & \textbf{79.7} & 46.1          & \textbf{75.9} & \textbf{72.1} & \textbf{77.5} & \textbf{76.6} & 54.3          & \textbf{86.1} & \textbf{78.8} \\
2 & NOT CIIR & 71.8          & 70.7          & 80            & 81            & \textbf{75.8} & 71.2          & 68.3          & 68.1          & 77.5          & \textbf{47.5} & 72.3          & 72            & 74.1          & 72.8          & 53.5          & 85.3          & 78.7          \\
3 & bott     & 71.6          & 70.3          & 82            & 79.8          & 70.4          & 70.6          & 67.5          & 69            & 76.5          & 44.4          & 75.2          & 71            & 75.7          & 75.6          & \textbf{55.5} & 85.2          & 77.5          \\
3 & mia      & 71.6          & 70            & \textbf{82.1} & 79.8          & 70.3          & 70.4          & 67.2          & 68.9          & 76.5          & 44.4          & 75.1          & 71            & 75.8          & 75.5          & 55.3          & 85.3          & 77.6          \\
3 & ew       & 71.6          & 70            & 82            & 79.9          & 70.3          & 70.4          & 67.2          & 68.9          & 76.5          & 44.4          & 75.1          & 71            & 75.7          & 75.6          & 55.2          & 85.3          & 77.6   \\
\bottomrule
\end{tabular}
}
\end{table}

Our massive first-stage hybrid and then reranking hybrid has allowed us to achieve the best result in test-a by a very slim margin of 0.2 (81.2 vs 81.0) and a slightly larger margin on test-b (1.5, 73.3 vs 71.8). Actually, our margin on the test-b is larger than the margin between the 2nd and 11th places, which means that the competition was pretty close with a lot of teams clustered together. Overall there is not one specific language where we improve a lot, our largest margins are French (+2.7) and Hindi (+2.2). However, there are no languages that we perform badly, as we are always no farther than 1.4 points (also smaller than our ``victory margin'') away from the best of that language (Indonesian). 

\section{The surprise languages}
\label{sec:surprise}

The second track of MIRACL was the surprise languages track. The idea was that 2 new languages would be added to the overall dataset, but this time without any training data and without having the time to prepare for them. In other words, the goal was to really test the capabilities of the systems to adapt to other languages. These two languages were Yoruba and German, with Yoruba having the lowest amount of documents of all languages, while German has one of the largest corpora of the entire 18 languages.

\subsection{First stage}

Given all the models that we had already trained we wanted to avoid training new models. However, it was kind of a given to at least try for a monolingual SPLADE in German as there was data available in mMARCO~\cite{mmarco}. Results for the development set are made available in Table~\ref{tab:all_first_surprise}. Our mSPLADE models did a very poor job on German, which was kinda expected as the models with the larger tokenizer did not learn well and the one with the sTok did not know about words in the surprise languages. Nevertheless, mSPLADE-sTok was actually a part of our final ensemble for Yoruba, which was quite surprising. For German mContriever had the better performance, while for Yoruba what really helped us was translating to English and then searching in English, with an nDCG@10 of 82.7. Hybrids were actually penalized a lot because of the poor performance of the mSPLADE models in German, but the ``Best'' hybrid still got an advantage over all the others. Finally, we got a 100\% recall@10 for Yoruba, which is something that we had not seen in any of the other languages.

\begin{table}[ht]
\centering
\caption{Comparison of all our first stage models on the surprise dev set.}
\label{tab:all_first_surprise}
\adjustbox{max width=\columnwidth}{%
\begin{tabular}{cc|cc|c}
\toprule
\# & model                    & German & Yoruba & Average \\
\midrule
\multicolumn{5}{c}{Models} \\
\midrule
a & BM25                & 22.6    & 40.6 & 31.6      \\
b & mDPR                & 48.9    & 44.4 & 46.7      \\
c & T-SPLADE            & 43.0    & \textbf{82.7} & 62.9      \\
d & SPLADE-mono         & n/a     & 39.8 & n/a \\
e & mSPLADE-small       & n/a     & n/a  & n/a \\
f & mSPLADE-large       & 9.4     & 47.5 & 28.5      \\
g & mSPLADE-sTok        & 9.7     & 51.3 & 30.5     \\
h & mContriever         & \textbf{51.8}    & 55.1 & 53.5      \\
i & mColBERT            & 49.7    & 36.6 & 43.2      \\
\midrule
\multicolumn{5}{c}{Hybrids} \\
\midrule
0 & a+b                            & 54.9    & 61.6 & 58.3     \\
1 & a+b+c                          & 64.8    & 84.5 & 74.7     \\
2 & a+b+c+d+e+f+g                  & 61.2    & 84.6 & 72.9     \\
3 & a+b+c+d+e+f+g+h+i              & 63.7    & 84.7 & 74.2     \\
Best & $\max$(a+b+c+d+e+f+g+h+i)   & \textbf{67.4} & \textbf{88.3} & \textbf{77.9}        \\
Simple & a+c+f+g                   & 56.1    & 86.1 & 71.1    \\
\bottomrule
\end{tabular}
}
\end{table}

\subsection{Rerankers}

For the reranking we actually used a few different rerankers compared to the known languages track. \footnote{This was mostly due to desperation when we saw everyone overtaking us on the dev set. These rerankers were then tested on known languages and they did not improve the results.} For Yoruba, we add three models a) the monoT0pp over the translated corpus we had used in English on the known languages track; b) RankT5 using ByT5~\cite{byt5}\footnote{\url{https://huggingface.co/google/byt5-xl}} as its pretrained language model and trained solely on mMARCO; and c) A cross encoding reranker using Deberta~\cite{deberta}\footnote{\url{https://huggingface.co/microsoft/mdeberta-v3-base}} as its pretrained language model. Results are presented in Table~\ref{tab:final_reranking_surprise}. The first thing we notice is that for the surprise languages, the rerankers actually improved more than for the known languages, showing that they are more robust to the change of language. 

\begin{table}[ht]
\centering
\caption{Full reranking results over ``Best'' first stage hybrid on the surprise languages.}
\label{tab:final_reranking_surprise}
\adjustbox{max width=\columnwidth}{%
\begin{tabular}{cc|ccc}
\toprule
\# & model                   & German & Yoruba & Average \\
\midrule
\multicolumn{5}{c}{First stage hybrids} \\
\midrule
F & Best & 67.4 & 88.3 & 77.9 \\
\midrule
\multicolumn{5}{c}{Rerankers} \\
\midrule
$\alpha$   & mono mT5-13b & 70.6    & \textbf{94.3} & \textbf{82.5}   \\
$\beta$    & Rank mT5-13b & \textbf{71.5}    & 92.7 & 82.1   \\
$\gamma$   & Rank mT0-xl  & 70.1    & 93.1 & 81.6   \\
$\delta$   & Rank mT0-xxl & 69.8    & 91.2 & 80.5  \\
$\epsilon$ & Rank Bloomz  & 55.3    & 76.5 & 65.9   \\
$\zeta$    & Translate + Rank T5 & 69.1 & 90.1 & 79.6   \\
$\eta$     & Translate + mono T0pp & 65.7 & 90.6 & 78.2 \\
$\theta$   & mono ByT5 & 68.5 & 92.9 & 80.7 \\
$\lambda$  & DeBerta & 65.2 & 80.7 & 73.0 \\
\midrule
\multicolumn{5}{c}{Reranking hybrids} \\
\midrule
Simple & max(F + $\alpha$ + $\beta$) & 74.2 & 93.6 & 83.9     \\
Best & max(F + $\alpha$ + $\beta$ + $\gamma$ + $\delta$ + $\epsilon$ + $\zeta$+ $\eta$ + $\theta$ + $\lambda$) & \textbf{74.4} & \textbf{94.8} & \textbf{84.6}    \\
\midrule
\bottomrule
\end{tabular}
}
\end{table}

\subsection{Comparison against other teams}

\paragraph{Dev Leaderboard}

So the first thing that we want to discuss when comparing to other teams is the leaderboard during the dev phase. We got really nervous when we saw that a lot of other teams were overtaking us, and we started adding a lot of models (as seen in the previous subsection). Final leaderboard results are made available in Table~\ref{tab:dev_surprise}. Our main analysis a-posteriori is that as there was no training data for both languages, it led to other teams overfitting to the dev set, but we don't have any idea if that is actually the case.

\begin{table}[ht]
\centering
\caption{Leaderboard for the development set on the surprise languages.}
\label{tab:dev_surprise}
\adjustbox{max width=\columnwidth}{%
\begin{tabular}{cc|cc|c}
\toprule
\# & team & German & Yoruba & Average \\
\midrule
\multicolumn{5}{c}{Models} \\
\midrule
1 & dm                  & \textbf{91.2}    & 85.3 & \textbf{88.3}      \\
2 & yesbody             & 78.0    & 94.5 & 86.2      \\
3 & bott                & 78.6    & 93.2 & 85.9      \\
4 & eternal             & 78.4    & 93.1 & 85.8 \\
5 & mia                 & 78.4    & 92.8 & 85.6 \\
6 & ew                  & 78.4    & 92.8 & 85.6      \\
7 & NOT CIIR            & 77.7    & 93.3 & 85.5     \\
8 & ale                 & 78.4    & 92.0 & 85.2      \\
9 & nle \textbf{(ours)} & 74.4    & \textbf{94.8} & 84.6      \\
\bottomrule
\end{tabular}
}
\end{table}

\paragraph{Test-B Leaderboard}

Much to our surprise the results of Test-B were completely different from what we had seen on the development set and are presented in Table~\ref{tab:test_b_surprise}. Our reranking ensemble which was the weakest of the top 10 on the development actually was the strongest on the test-b, while we still kept our advantage in the Yoruba language. We probably got lucky that our models worked pretty well without needing to retrain specifically on the surprise languages and thus were more robust to the change from the surprise to the test-b. 

\begin{table}[ht]
\centering
\caption{Leaderboard for the test-b set on the surprise languages.}
\label{tab:test_b_surprise}
\adjustbox{max width=\columnwidth}{%
\begin{tabular}{cc|cc|c}
\toprule
\# (\# dev) & team & German & Yoruba & Average \\
\midrule
\multicolumn{5}{c}{Models} \\
\midrule
1 (9) & nle \textbf{(ours)} & \textbf{76.6}    & \textbf{95.0} & \textbf{85.8}      \\
2 (3) & bott                & 73.7    & 93.2 & 83.5      \\
3 (5) & mia                 & 73.6    & 93.2 & 83.4      \\
4 (6) & ew                  & 73.5    & 93.2 & 83.4 \\
\bottomrule
\end{tabular}
}
\end{table}

\section{Query analysis and labeling}

Looking back, we may have made things actually harder for us. If one considers that MIRACL labeling was generated mostly based on BM25/mDPR/mColBERT\footnote{which is different from our mColBERT}, which are weaker than the ones considered in this work, we could be just increasing the amount of false negatives. This is what we aim to study in this section. Note that we do not want to bash the annotation of MIRACL, a dataset with so many queries and so much annotation is not only essential to the community but very costly (both in monetary value and amount of work), our goal here is to look into if we made it harder for ourselves or if spending so much time on first stage retrievers was worth it.

Just as a preliminary analysis, let us consider the Judged@ metric. As there's an average of 10 documents judged per query, one would expect that Judged@10 increases a lot when nDCG@10 increases. However this is not actually what we see, for example, Hybrid 0 (nDCG@10 57.8) has an average Judged@10 of 50.1, while our best run actually reduces it to 48.5\%. While this is probably due to the fact that we learn to not retrieve the known negatives, it is still counter-intuitive to what one would expect (for example recall@20 increased from 83.0\% to 94.0\%). This could mean that there are potential false negatives that we found, that would not have been found with BM25 or mDPR, making it harder for the rerankers as not only do they need to learn to properly rank, but also to avoid the labeling bias. 

However, actually looking at the data from French\footnote{We could also have looked into English, but much more time would be needed as there are more than 100 queries where the top1 document is not judged.} actually showed the opposite: i.e. There are more false negatives on our top1 that were also present on the top100 of Hybrid 0 than the opposite. The French dev set is composed of 343 queries, of which 217 are judged as positive for our submitted model, 93 are known negatives and only 33 (slightly less than 10\%) are unjudged. Of those 33, 5 were not on the top200 of Hybrid 0 and of those 5 only 2 (40\%) is actually a false positive and 1 of the others we do not agree with the positive passage. On the other hand, out of the other 28, 15 (53\%) are false negatives\footnote{Note that this is actually 4\% of the total queries of the devset and thus not that relevant} and another 4 (making 19/28 or around 67\%) would actually be positives if the date/important information was not removed in the crawling process, for example, consider the following query and its translation\footnote{translation made by a fluent but non-native speaker of both languages}: \begin{quote}
    1260128\#0 | ``Quand les premiers Oscars ont-ils eu lieu?'' | ``When did the first Academy Award (Oscars) ceremony happened?''
\end{quote}
The document found by our system was (Title - Document):
\begin{quote}
11044\#19 | ``Oscars du cinéma - Initialement organisée sous forme de banquet, la première cérémonie, présidée par Douglas Fairbanks, a lieu le à \dots'' | ``Academy Awards (Oscars) - The first ceremony, initially organized as a private dinner function and presided by Douglas Fairbanks, happened on at\dots''
\end{quote}

Note that the actual dates and place are missing. What we actually believe happened is as some dates are actually links they were removed by the crawling process. 

\paragraph{Summary of analysis:} Our analysis actually shows that the labeling of the dataset is a smaller problem than we warranted at first, but has led us to find some possible concerns on the Wikipedia crawl missing important parts of the documents. It also shows that the models tend to highly rank a document for a ``date'' based question, even if no date is actually present in the document, making it so that the model kinda knows the format of a response for a date based question, but does not care about the answer itself, more about its format.

\section{Conclusion}
For the WSDM 23 CUP - MIRACL, we used all the knowledge we had from previous participations at TREC in order to generate a large ensemble of models. While there's probably not much difference between the full ensemble and a smaller smarter subset of models we were actually able to draw from the knowledge embedded in each model, leading to winning both tracks (known languages and surprise languages). As a first start for the MIRACL dataset, we found that this competition was really fruitful and we hope to get a better understanding of what made our runs work so that we can make them more efficient and need less computation in the future. We also hope that this experiment report is useful for the readers and are more than happy to discuss further the whole process of transferring the English-based methods to a multilingual setting.

\section*{Acknowledgement}

We would like to thank the organizers of the WSDM 23 CUP - MIRACL for all their hard work in making the competition not only take place, but run smoothly. They did a mostly thankless job of getting more than 700k dedicated annotations for the multilingual dataset used in this competition, which will for sure foster development in this much-needed subject.

\bibliographystyle{abbrv}
\bibliography{sample-base}





\end{document}